\documentclass[runningheads]{comsis2}

\def\journalissue{Computer Science and Information Systems 2021, 18(1): 303--321}
\def\paperidnum{https://doi.org/10.2298/CSIS201109059Z}
\usepackage[pagewise]{lineno}

\usepackage{hyperref}
\usepackage{verbatim}
\usepackage{graphicx}
\usepackage{subfigure}
\usepackage[justification=centering]{caption}
\newcommand{\tc}{\textsuperscript}
\usepackage{array}
\usepackage{booktabs}

\title{Metaphor Research in the 21st Century: A Bibliographic Analysis}

\footnotetext[1]{Corresponding author: Feng Xia; e-mail: f.xia@ieee.org}

\titlerunning{Metaphor Research in the 21st Century}
\author{Dongyu Zhang\inst{1} \and Minghao Zhang\inst{1} \and
        Ciyuan Peng\inst{2} \and Jason J. Jung\inst{2} \and Feng Xia\inst{3}}

\authorrunning{Zhang et al.}
\institute{School of Software, Dalian University of Technology, Dalian 116620, China\\
  \and
  Department of Computer Engineering, Chung-Ang University, Seoul 156-756, Korea\\
  \and
  School of Engineering, IT and Physical Sciences, Federation University Australia, Ballarat 3353, Australia\\
}

\begin{document}

\maketitle

\begin{abstract}
Metaphor is widely used in human communication. The cohort of scholars studying metaphor in various fields is continuously growing, but very few work has been done in bibliographical analysis of metaphor research. This paper examines the advancements in metaphor research from 2000 to 2017. Using data retrieved from Microsoft Academic Graph and Web of Science, this paper makes a macro analysis of metaphor research, and expounds the underlying patterns of its development. Taking into consideration sub-fields of metaphor research, the internal analysis of metaphor research is carried out from a micro perspective to reveal the evolution of research topics and the inherent relationships among them. This paper provides novel insights into the current state of the art of metaphor research as well as future trends in this field, which may spark new research interests in metaphor from both linguistic and interdisciplinary perspectives.

\vspace{6pt}\textbf{Keywords:} metaphor, literature analysis, statistical analysis, scholarly big data.
\end{abstract}

\section{Introduction}
\label{intro}
Metaphor is an indispensable part of human communication. According to empirical studies, every three sentences in natural language uses a metaphor \cite{richards1965philosophy,steen2010method}. It is not only a universal linguistic phenomenon but also a means for people to understand and cognize \cite{lakoff1980metaphors}. Humans frequently use one concept in metaphors to describe another concept for reasoning. For instance, in the metaphorical utterance: `experience is a treasure,' we use `treasure' to describe `experience' to emphasize that `experience' can be valuable. A metaphor has been viewed as a mapping system that conceptualizes one domain (target) in terms of another (source)~\cite{lakoff1980metaphors}. In particular, along with the rapid explosion of social media applications such as Facebook and Twitter, metaphorical texts and information have increased dramatically. It seems to be very common for Internet users to use vivid and colorful metaphorical expressions on social media on a variety of topics including, products, services, public events, tidbits of their life, etc.

An increasing number of researchers have studied metaphor from different perspectives in fields like linguistics \cite{tendahl2008complementary,semino2017corpus,Rebecca2020,Rong2020}, psychology~\cite{norris2008community,kopp2013metaphor,MARKOWITZ2020100816,power2020}, neuroscience~\cite{aziz2008embodied,hellberg2018funny,garson2019,gulli2019}, management~\cite{woodside2018organizational,RINCONRUIZ2019100924,Fortin2020,BELHASSEN2020100228}, and computer science~\cite{tsvetkov2014metaphor,parde2018exploring,wu2018neural,hall2020metaphor}. Since metaphor research has been developing dramatically, it is necessary to review the current situation, the development and trends of metaphor research, as well as studying how metaphor research has evolved through time. This may make contributions to some novel and interesting studies of metaphors from both linguistic and interdisciplinary perspectives as well as exploring the related underlying mechanism. Previous studies have shown that quantitative analysis can explain the nature of a particular discipline or field and changes in research focus over time \cite{meyer2009development,correia2018scientometric}. Researchers can use some information platforms, such as AMiner~\cite{Wan2019}, Google Scholar \cite{Delgado2019}, Microsoft Academic Services \cite{Kuansan00021}, and many other scientific online systems \cite{2020arXiv200510732V}. These information platforms contain useful data, including but not limited to authors, papers, and references, and they can carry out statistical analysis. So far, based on the above academic systems, a large number of related works have applied quantitative analysis techniques in scientometrics. \cite{kong2018human} used bibliographic analysis to summarize human interactions. \cite{kaushal2018emerging} reviewed various studies using online social networks to identify personality, as reported in the literature. \cite{correia2018scientometric} made a quantitative assessment of mapping the intellectual structure and development of computer-supported cooperative work. \cite{liu2018artificial} used complex network topology to study the evolution of artificial intelligence. Also, \cite{sun2017discovering} made contributions to the research in the field of transportation.

Numerous theories and technologies of literature analysis based on big scholarly data have been proposed \cite{wang2017shifu,shifu2,xia2017big,KONG201986}. However, so far, few people have collected bibliometrics data to analyze metaphor quantitatively and to comprehend its internal structure as well as evolution. To fill this gap, in this paper, we carry out a bibliometric analysis of the development of metaphor research in the early 21\tc{st} century, based on the following four aspects. First, we analyze the development of metaphor research by counting the increment of the number of publications over time.  Second, we emphasize influence and citation patterns to distinguish the behavioral dynamics of citation. Third, we try to quantify milestones during this period through identification of the characteristics of influential papers, researchers, and institutions. Finally, we explore the internal structure of metaphor research by analyzing the evolution of themes and mutual attraction.

\begin{figure}[h]
	\centering
	\includegraphics[width=0.7\textwidth]{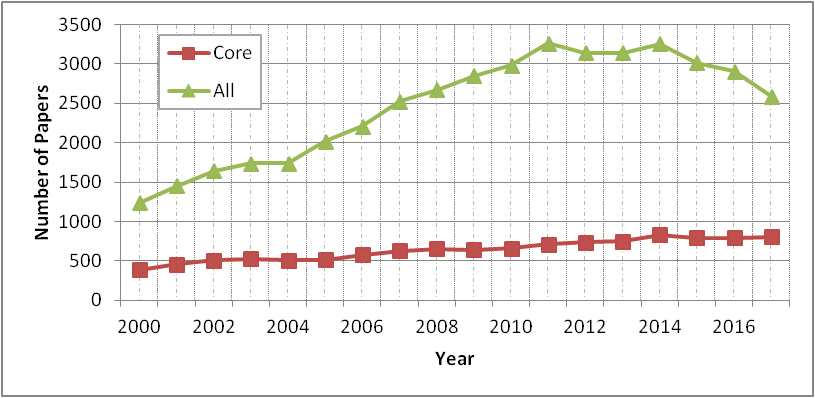}
	\caption{Changes in the number of papers in Metaphor (every year) since 2000.}
	\label{figure1}
\end{figure}

\begin{figure}[h]
	\centering
	\includegraphics[width=0.7\textwidth]{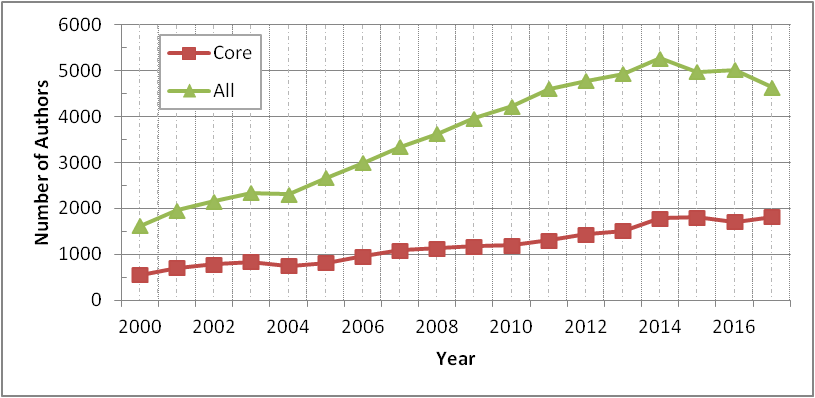}
	\caption{The number of authors every years.}
	\label{figure2}
\end{figure}

The scholarly dataset we use in our study consists of 11,564 papers from the Web of Science and 44,586 papers from Microsoft Academic Graph (MAG). The rest of the paper is organized as follows. Section \ref{sec2} provides the methodologies and models we use for our analysis. Section \ref{sec3} introduces the experimental results we obtained from our literature analysis. Section \ref{sec4} concludes the paper and provides some directions for the future.

\section{Methods}
\label{sec2}
In this section, we first introduce the data set we use to analyze metaphors: core data sets and extended data sets. Then we introduce several indicators for measuring the importance of authors and institutions in the field of metaphor research and their calculation methods. Finally, we introduce the division of the field of metaphor research.

\subsection{Datasets}
For conducting experiment, we employ MAG data set (\url{http://research.microsoft.com/en-us/projects/mag/}) ---a widely used and one of the best databases for empirical research in scientometrics and citation analysis \cite{sinha2015overview,Kuansan00021}. Hence, to investigate the current state of metaphor research, we extract the papers from the MAG data set, which contains six entities: affiliations, authors, conferences, fields of study, journals, and papers. The new MAG data set contains new relationships in the field of study with papers. First, we limit the publication time of the articles to 2000 and beyond. Then, from these papers, we select articles that comprise at least one of the following six words in their title or abstract: metaphor, metaphorical, metaphorically, Metaphor, Metaphorical, or Metaphorically. We use all the extracted papers as our extended data set containing 44,586 articles, of which 1,872 are conference papers. Because the number of conference papers was inadequate, we do not consider its particularity, and we do not give it any special treatment.

Additionally, we found all the journals related to metaphors from the Web of Science database (\url{http://isiknowledge.com}), of which there are about a thousand. We extract these journals as a list of our core journals. Then, based on the list of core journals, we extract the articles published in the core journals from the papers of the extended data set as our core data set. It contains 11,564 articles.

\begin{figure}[h]
	\centering
	\includegraphics[width=0.7\textwidth]{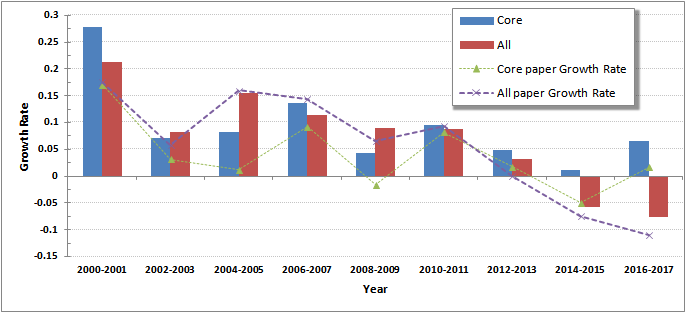}
	\caption{The growth rate of authors as well as total publications every two years.}
	\label{figure3}
\end{figure}

\begin{figure}[h]
	\centering
	\includegraphics[width=0.7\textwidth]{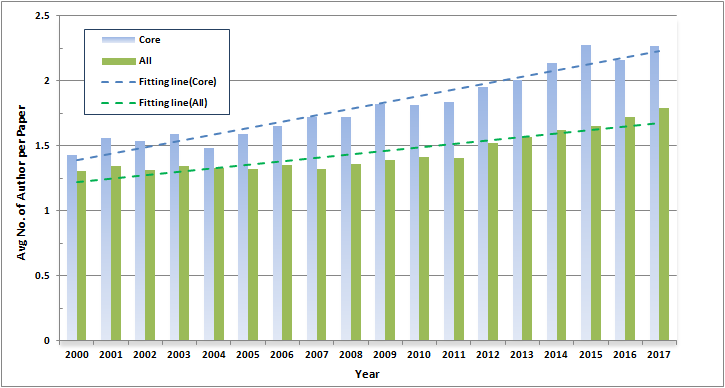}
	\caption{The average number of authors per paper.}
	\label{figure4}
\end{figure}

We use the same statistics and calculations for both the core data set and the extended data set.

\begin{table}[h]
	\caption{Ranking of papers based on the total number of citations received in2000-2017 in core dataset papers.}
	\label{tab:table1}
	\begin{tabular}{lp{8.5cm}p{1.3cm}<{\centering}p{2cm}<{\centering}}
		\toprule
		No.&Title&Citations & Published Year\\
		\midrule
		1 & Knowledge and organization: A social-practice perspective\cite{brown2001knowledge}& 2,044&2001\\
		2 & The network structure of social capital\cite{burt2000network}& 1,922&2000\\
		3 & Adaptive subgradient methods for online learning and stochastic optimization\cite{duchi2011adaptive}& 1,609&2011\\
		4 & From metaphor to measurement: Resilience of what to what?\cite{carpenter2001metaphor}& 1,348&2001\\
		5 & Community resilience as a metaphor, theory, set of capacities, and strategy for disaster readiness\cite{norris2008community}& 1,279&2008\\
		6 & Social and psychological resources and adaptation\cite{hobfoll2002social}& 1,234&2002\\
		7 & Relational frame theory: A post-Skinnerian account of human language and cognition\cite{barnes2001relational}& 1,185&2001\\
		8 & Modern social imaginaries\cite{taylor2002modern}& 1,150&2002\\
		9 & Where mathematics comes from: How the embodied mind brings mathematics into being\cite{lakoff2000mathematics}& 993&2000\\
		10 & The evolution of foresight: What is mental time travel and is it unique to humans?\cite{suddendorf2007evolution}& 800&2007\\
		11 & A thorough benchmark of density functional methods for general main group thermochemistry, kinetics, and noncovalent interactions\cite{goerigk2011thorough}& 724&2011\\
		12 & Self-control relies on glucose as a limited energy source:  Willpower is more than a metaphor\cite{gailliot2007self}& 723&2007\\
		13 & Scale-free networks provide a unifying framework for the emergence of cooperation\cite{santos2005scale}& 696&2005\\
		14 & The surveillant assemblage\cite{haggerty2000surveillant}& 692&2000\\
		15 & Metaphoric structuring: Understanding time through spatial metaphors\cite{boroditsky2000metaphoric}& 685&2000\\
		\bottomrule
	\end{tabular}
\end{table}

\subsection{Indicators and calculation methods}
We consider the following indicators to assess the relevance of authors as well as publications in this field.

\begin{figure}[h]
	\centering
	\includegraphics[width=0.7\textwidth]{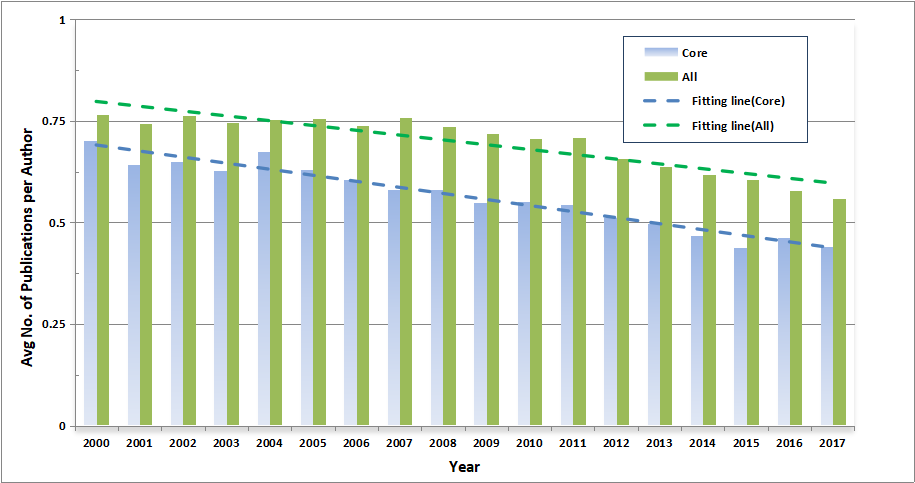}
	\caption{The average productivity of Metaphor scientists.}
	\label{figure5}
\end{figure}

\begin{figure}[h]
	\centering
	\includegraphics[width=0.7\textwidth]{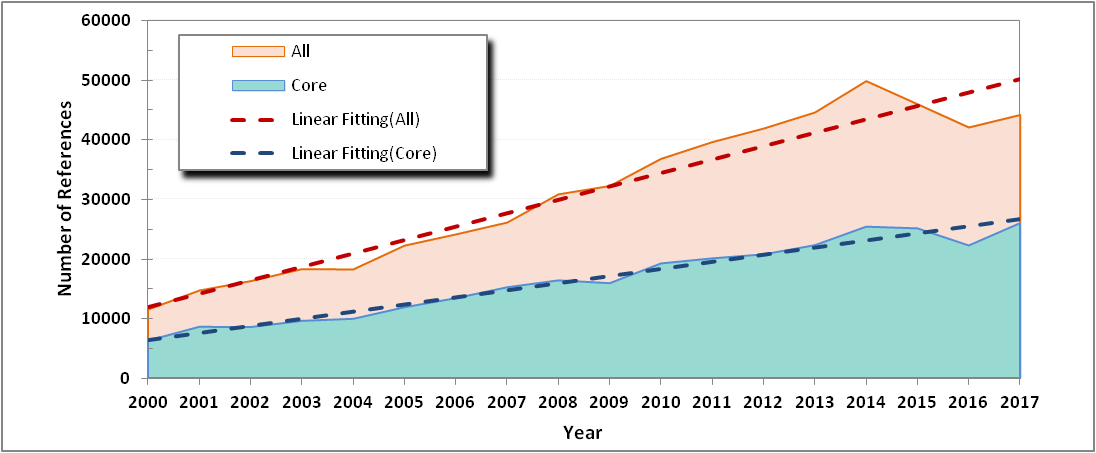}
	\caption{Changes in references.}
	\label{figure6}
\end{figure}

\begin{itemize}
	\item {\verb|Measuring research output by measurement|}: We assume that the core data set or the extended data set is $P$, and we use statistical methods to calculate the total number of articles in the data set denoted as $\vert P \vert$, total number of authors $\sum_{p \in P} \vert A_p \vert $, total number of citations $\sum_{p \in P} \vert Ci_p \vert $, and total number of references $\sum_{p \in P} \vert R_p \vert $. We then calculate the average number of authors per paper $\frac {\sum_{p \in P} \vert A_p \vert}{\vert P \vert}$, the average number of citations per paper $\frac {\sum_{p \in P} \vert Ci_p \vert}{\vert P \vert}$, the average number of references per paper $\frac {\sum_{p \in P} \vert R_p \vert}{\vert P \vert}$, and the average number of papers per author $\frac {\vert P \vert}{\sum_{p \in P} \vert A_p \vert}$ ($\vert A_p \vert$ represents the total number of authors of the paper, $\vert Ci_p \vert$ represents the total number of citations of the paper, and $\vert R_p \vert$ represents the total number of references of the paper).
	
	\begin{figure}[h]
		\centering
		\includegraphics[width=0.7\textwidth]{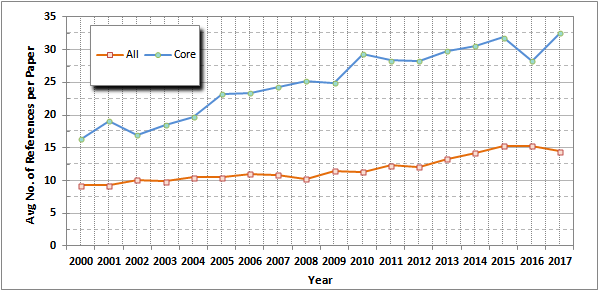}
		\caption{Average number of references per paper.}
		\label{figure7}
	\end{figure}
	
	\item{\verb|self-citation rate|}: In addition, to reflect the dynamics of the researcher's reference behavior, we use the most rigorous self-guided definition as our evaluation height, that is, if both referenced papers have at least one mutual  author, then there are two references between these references. The paper is self-cited by the author. It can be computed as $\frac {\sum_{r \in R} \vert A_r \vert}{\vert R \vert}$, where $\vert R \vert$ is the total number of references of the paper and $\vert A_r \vert$ is the number of author self-citation. Similarly, a self-journal (conference) is when the paper and one or more of its references are published in the same journal (conference). This can be computed as $\frac {\sum_{r \in R} \vert J_r \vert}{\vert J_r \vert}$,  where $\vert R \vert$ is the total number of references in the paper and $\vert J_r \vert$ is the number of journal (conference) self-citations. Self-affiliation is when the paper and one or more of its references come from same affiliation. This can be computed as $\frac {\sum_{r \in R} \vert {Aff}_r \vert}{\vert R \vert}$, where $\vert R \vert$ is the total number of references in the paper and $\vert J_r \vert$ is the number of journal (conference) self-citations.
\end{itemize}

\begin{figure}[h]
	\centering
	\includegraphics[width=0.7\textwidth]{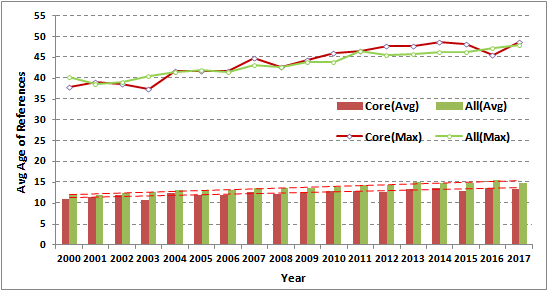}
	\caption{The average age differences between the cited paper and the citing paper.}
	\label{figure8}
\end{figure}

\subsection{The inner structure of metaphor}

\begin{itemize}
	\item {\verb|Topic exploration|}: The study of metaphor is not an independent discipline, but an interdisciplinary science. The MAG data set constructs the domain into a forest of six-layered tree structures. The new MAG data set also contains new relationships in the field of study.  Therefore, we can easily divide the topic of metaphor research. In the end, we choose the root node of each tree as the topic of metaphor research. The core data set and the extended data set have the same 19 topics: psychology, sociology, computer science, economics, medicine, biology, mathematics, philosophy, engineering, business, history, physics, political science, chemistry, geography, geology, art, environmental science, and materials science. We also select the top 50 secondary fields with the largest number of articles for our subsequent analysis.
	
	\item{\verb|The relevance of the topics|}: To investigate the relevance of the topics further, given the two topics $A$ and $B$, we calculate the probability of $B$ occurring given $A$’s occurrence as follows.
	
	First, we compute the probability of Topics $A$ and $B$’s occurrence as $P_A=\frac {N_A}{N}$ and $P_B=\frac {N_B}{N}$, where $\vert N_A \vert$ and $\vert N_B \vert$ represent the total number of papers belong to Topic $A$ and Topic $B$, respectively. $\vert N \vert$ is the total number of papers.	
	
	Second, we calculate the probability of Topics $A$ and $B$ appearing simultaneously as $P_{AB}=\frac {N_{AB}}{N}$, where $\vert N_{AB} \vert$ is the number of articles simultaneously belongs to Topic $A$ and Topic $B$, respectively.
	
	Finally, we obtain the probability that $A$ appears under the condition that $B$ appears by $P(A \vert B) = \frac {P_{AB}}{P_B}$.
	
	Using the above method, we calculate the probability relationships between the top 50 secondary fields containing the largest number of articles for later analysis work.
	
	\item{\verb|Proportion of the topic in different years|}: To observe the evolution of the topic over time, we use $\theta_k^{[t]}$ \cite{liu2018artificial} to represent the proportion of the topic $k$ at $t$ year. It can be seen that $\theta$ is the average topic distribution in all articles. This indicator allows us to quantify the importance of topics over a specific period. We compute the indicator in the root field.
	
	\begin{figure}[h]
		\centering
		\includegraphics[width=0.7\textwidth]{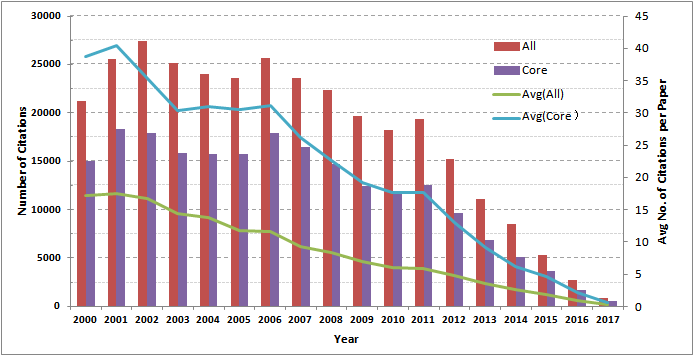}
		\caption{The number of citations per year and the average number of papers cited.}
		\label{figure9}
	\end{figure}
	
	\item{\verb|Popular topics|}: To measure the trend of a field over time,  we calculate the variety index between two periods with $r_k=\frac {\sum_{t=2010}^{2017} \theta_k^{[t]}}{\sum_{t=2000}^{2009} \theta_k^{[t]}}$. We compute the indicator in the root field. When $r_k > 1$, this field is more popular in 2010-2017 than in 2000-2009. While $r_k < 1$, the research in this field reduced in 2010-2017. Further, when $r_k = 1$, there is no change.
	
	\item{\verb|Network of topic co-presence|}: Article co-citation analysis is often used to identify developments in the field of research by exploring common citation relationships between references as a basis for assessing and planning scientific and technical research. In a visual network map, the lines in the document co-citation network represent the frequency with which other publications in the same data set refer to both publications. Based on the similarities of the research, the network can be divided into different groups. \cite{liu2018artificial} conducted an experiment of joint citation analysis to reveal the evolution of the field of social simulation. Following their steps, we use this method to build a topic coexistence network to discover the interconnection patterns between them. Based on the correlations of the subjects $P_A$, $P_B$, and $P_{AB}$, which we compute before, we calculate the coexistence coefficient $co=\frac {P_{AB}^2}{min(P_A,P_B)*mean(P_A,P_B)}$. Therefore, we choose themes with $co(A, B) > 0.1$ to build coexistence. We calculate the index between the top 50 secondary fields containing the largest number of articles for later analysis work.
\end{itemize}

\begin{table}
	\caption{Ranking of papers based on the total number of citations received in 2000-2017 in all dataset papers.}
	\label{tab:tabel2}
	\begin{tabular}{lp{8.5cm}p{1.3cm}<{\centering}p{2cm}<{\centering}}
		\toprule
		No.&Title&Citations & Published Year\\
		\midrule
		1 & Knowledge and organization: A social-practice perspective \cite{brown2001knowledge}& 2,044&2001\\
		2 & The network structure of social capital\cite{burt2000network}& 1,922&2000\\
		3 & Adaptive subgradient methods for online learning and stochastic optimization\cite{duchi2011adaptive}& 1,609&2011\\
		4 & From metaphor to measurement: Resilience of what to what?\cite{carpenter2001metaphor}& 1,348&2001\\
		5 & Community resilience as a metaphor, theory, set of capacities, and strategy for disaster readiness\cite{norris2008community}& 1,279&2008\\
		6 & Social and psychological resources and adaptation\cite{hobfoll2002social}& 1,234&2002\\
		7 & Relational frame theory: A post-Skinnerian account of human language and cognition\cite{barnes2001relational}& 1,185&2001\\
		8 & Modern social imaginaries\cite{taylor2002modern}& 1,150&2002\\
		9 & Negotiation as a metaphor for distributed problem solving\cite{Davis2003Negotiation}& 1,092&2003\\
		10 & Where mathematics comes from: How the embodied mind brings mathematics into being\cite{lakoff2000mathematics}& 993&2000\\
		11 & Animation: Can it facilitate?\cite{betrancourt2002animation}& 928&2002\\
		12 & Model-driven data acquisition in sensor networks\cite{deshpande2004model}& 848&2004\\
		13 & The evolution of foresight: What is mental time travel and is it unique to humans? \cite{suddendorf2007evolution}& 800&2007\\
		14 & A thorough benchmark of density functional methods for general main group thermochemistry, kinetics, and noncovalent interactions. \cite{goerigk2011thorough}& 724&2011\\
		15 & Self-control relies on glucose as a limited energy source:  Willpower is more than a metaphor \cite{gailliot2007self}& 723&2007\\
		\bottomrule
	\end{tabular}
\end{table}

\section{Results}
\label{sec3}
\subsection{Evolution of metaphor}

As Fig. \ref{figure1} shows, in the whole development process of metaphor research, the number of papers on metaphor published every year continues to increase in both the core data set and the whole data set. This finding shows that metaphor research has become more and more popular in recent years. Could this be the result of an increase in the number of researchers? To verify this conjecture, as shown in Fig. \ref{figure2}, we analyze the number of authors in the data set, and we found that the growth rate of authors has the same trend as the number of papers, but it is slightly higher (Fig. \ref{figure3}). We conclude that the increase in the number of authors might stimulate an increase in the number of metaphorical papers.

\begin{figure}[h]
	\includegraphics[width=\linewidth]{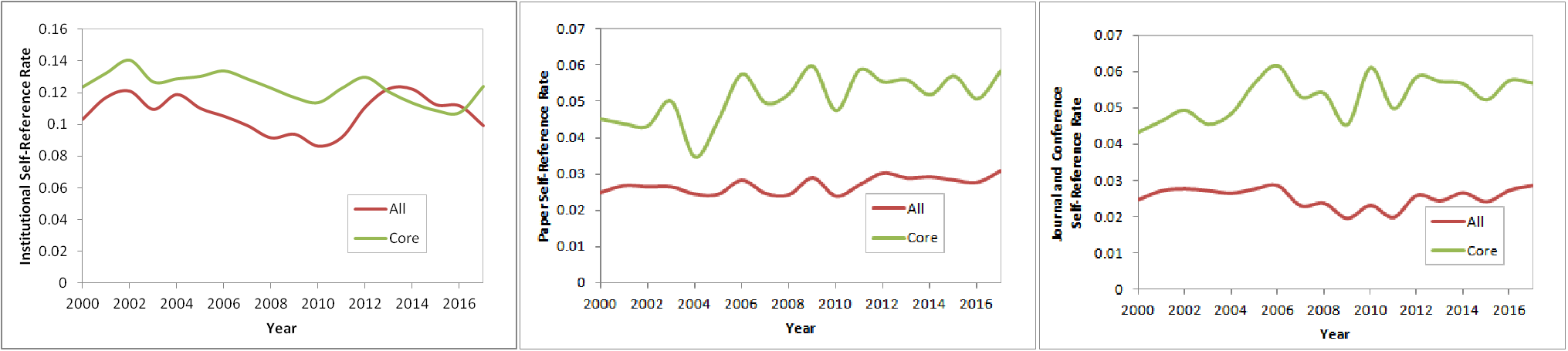}
	\caption{Institutional self-citation rate, Paper self-citation rate, and Conference and journal self-citation rate.}
	\label{figure10}
\end{figure}

Since some conferences are held every two years, the number of papers and the overall results are affected. The primary purpose of conferences is to provide opportunities for scientists to communicate and to understand what others are doing. They can publish their research results as soon as possible, which is very important for timely subjects. Journal papers, by contrast, have a longer review cycle, which can lead to fluctuations in growth rates. Most of the data in our statistics come from journal papers, and a small part come from conference papers. Although we put the two types of papers together for statistical analysis, to explain the development of this discipline better, we analyze the growth rate by using the data from every two years as a unit.

Besides, Fig. \ref{figure4} plots the average number of authors per paper over time, and a clear upward trend can be seen from the fit curve, indicating that collaborative papers are becoming more common. We also observe that the average number of publications per author declined over time (Fig. \ref{figure5}), indicating that average productivity was weakening during this period.

\begin{table}
	\caption{Ranking of authors based on the average number of citations per paper during 2000-2017 (Core dataset Author).}
	\label{tab:tabel3}
	\begin{tabular}{lp{3cm}p{4.5cm}p{1.5cm}p{1cm}p{1.5cm}}
		\toprule
		No.&Name&Organization&Citation&Paper&Citations per Paper\\
		\midrule
		1  & John Seely Brown    & PARC                                                         & 2044 & 1 & 2044 \\
		2  & Paul Duguid         & University of California, Berkeley                           & 2044 & 1 & 2044 \\
		3  & John C. Duchi       & Stanford University                                          & 1609 & 1 & 1609 \\
		4  & Elad Hazan          & Princeton University                                         & 1609 & 1 & 1609 \\
		5  & Yoram Singer        & Hebrew University of Jerusalem                               & 1609 & 1 & 1609 \\
		6  & Steve Carpenter     & University of Wisconsin-Madison                              & 1348 & 1 & 1348 \\
		7  & Nick Abel           & Commonwealth Scientific and Industrial Research Organisation & 1348 & 1 & 1348 \\
		8  & J. Marty Anderies   & Commonwealth Scientific and Industrial Research Organisation & 1348 & 1 & 1348 \\
		9  & Brian Walker        & Commonwealth Scientific and Industrial Research Organisation & 1348 & 1 & 1348 \\
		10 & Rose L. Pfefferbaum & Phoenix College                                              & 1279 & 1 & 1279 \\
		11 & Karen Fraser Wyche  & University of Oklahoma Health Sciences Center                & 1279 & 1 & 1279 \\
		12 & Betty Pfefferbaum   & University of Oklahoma Health Sciences Center                & 1279 & 1 & 1279 \\
		13 & Fran H. Norris      & Dartmouth College                                            & 1279 & 1 & 1279 \\
		14 & Susan P. Stevens    & Dartmouth College                                            & 1279 & 1 & 1279 \\
		15 & Stevan E. Hobfoll   & Rush University Medical Center                               & 1234 & 1 & 1234\\
		\bottomrule
	\end{tabular}
\end{table}

\subsection{Impact and citation analysis}

A dramatic increase in the number of references (Fig. \ref{figure6}) indicates that researchers are more focused on the work of others. The reason for this phenomenon may be the increase in the number of references per paper and the increase in the number of published papers (Fig. \ref{figure1}). From Fig. \ref{figure7}, we can see the change in the average number of references for each paper from 2000 to 2017. In the core data set, the average number of references per paper increased from 16 in 2000 to 32 in 2017. Papers in all data sets have the same trend (from 9 in 2000 to 15 in 2015).
Fig. \ref{figure8} shows the average and the maximum age difference between the cited paper and the citation. We can see that the age difference in the reference cited by researchers shows a prolonged and tortuous growth trend. The main reason for this phenomenon is that scholars cite papers in different ways: one is mainly referring to classic papers, and another refers to the latest papers. In 2012, \cite{krizhevsky2012imagenet} first used deep learning to classify high-resolution images, confirming that deep convolutional neural networks are superior to traditional machine learning techniques. More and more scholars are trying to keep abreast of the latest developments, which reduces the average age difference between citations and cited papers, and which restricts the impact of reference classic papers.		

\begin{figure}
	\centering
	\subfigure[Total dataset]{
		\centering
		\includegraphics[width=0.45\linewidth]{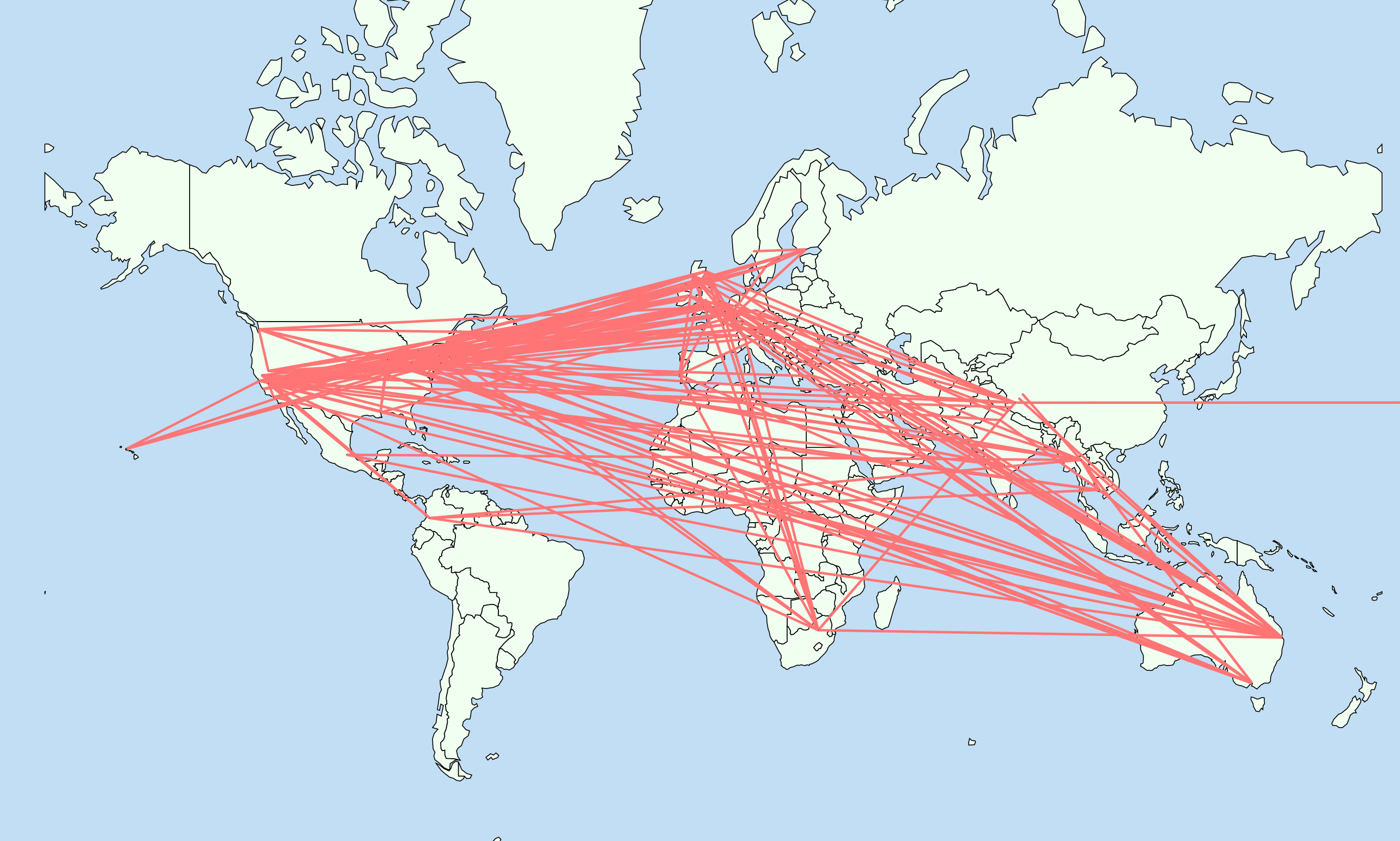}
	}%
	\subfigure[Core dataset]{
		\centering
		\includegraphics[width=0.45\linewidth]{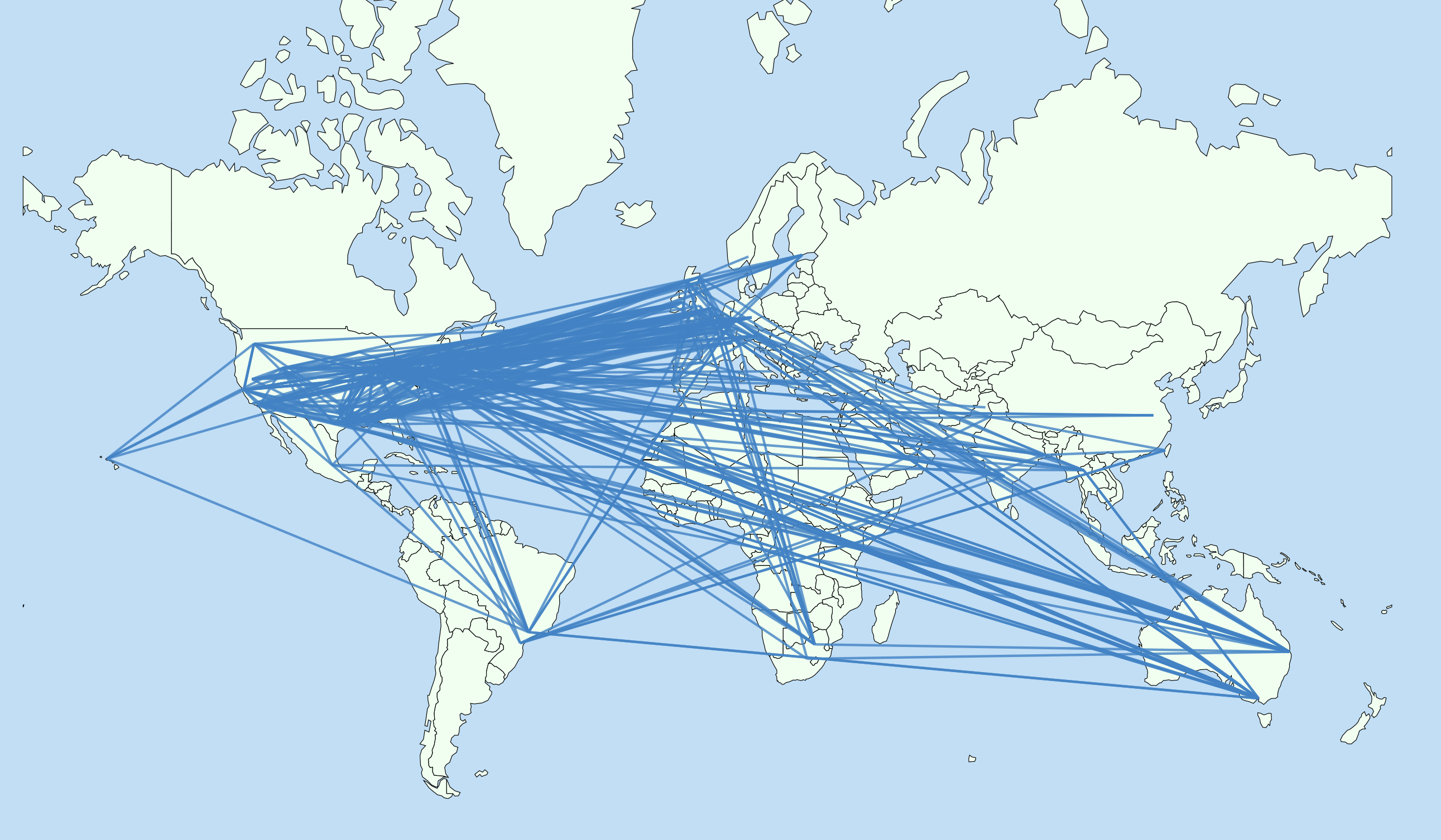}
	}%
	\caption{The overview of Metaphor citation relationships between 2000 and 2017. The lines represent the citation relationships among the top 50 most-cited institutions.}
	\label{fig:11}
\end{figure}
%
In general, the more recently published papers are, the fewer the people who read and cite them, the shorter the time of citations, and the lesser the impact. For example, if a paper published today is not known to anyone, it will not be cited. Therefore, the average number of citations per paper should decrease as publication time approaches. However, as shown in Figure \ref{figure9}, there are still years with increased citations, such as 2001, 2006, and 2011,  indicating that the papers published in those years are more influential than other years.	

\begin{table}
	\caption{Ranking of authors based on the average number of citations per paper during 2000-2017 (All dataset Author).}
	\label{tab:tabel4}
	\begin{tabular}{lp{3cm}p{4.5cm}p{1.5cm}p{1cm}p{1.5cm}}
		\toprule
		No.&Name&Organization&Citation&Paper&Citations per Paper\\
		\midrule
		1  & John Seely Brown    & PARC                                                         & 2044 & 1 & 2044 \\
		2  & Paul Duguid         & University of California, Berkeley                           & 2044 & 1 & 2044 \\
		3  & John C. Duchi       & Stanford University                                          & 1609 & 1 & 1609 \\
		4  & Elad Hazan          & Princeton University                                         & 1609 & 1 & 1609 \\
		5  & Yoram Singer        & Hebrew University of Jerusalem                               & 1609 & 1 & 1609 \\
		6  & Steve Carpenter     & University of Wisconsin-Madison                              & 1348 & 1 & 1348 \\
		7  & Nick Abel           & Commonwealth Scientific and Industrial Research Organisation & 1348 & 1 & 1348 \\
		8  & J. Marty Anderies   & Commonwealth Scientific and Industrial Research Organisation & 1348 & 1 & 1348 \\
		9  & Rose L. Pfefferbaum & Phoenix College                                              & 1279 & 1 & 1279 \\
		10 & Karen Fraser Wyche  & University of Oklahoma Health Sciences Center                & 1279 & 1 & 1279 \\
		11 & Betty Pfefferbaum   & University of Oklahoma Health Sciences Center                & 1279 & 1 & 1279 \\
		12 & Fran H. Norris      & Dartmouth College                                            & 1279 & 1 & 1279 \\
		13 & Susan P. Stevens    & Dartmouth College                                            & 1279 & 1 & 1279 \\
		14 & Stevan E. Hobfoll   & Rush University Medical Center                               & 1234 & 1 & 1234 \\
		15 & Bryan Roche         & Maynooth University                                          & 1185 & 1 & 1185\\
		\bottomrule
	\end{tabular}
\end{table}

\begin{figure}[h]
	\centering
	\subfigure[Total dataset]{
		\centering
		\includegraphics[width=0.45\textwidth]{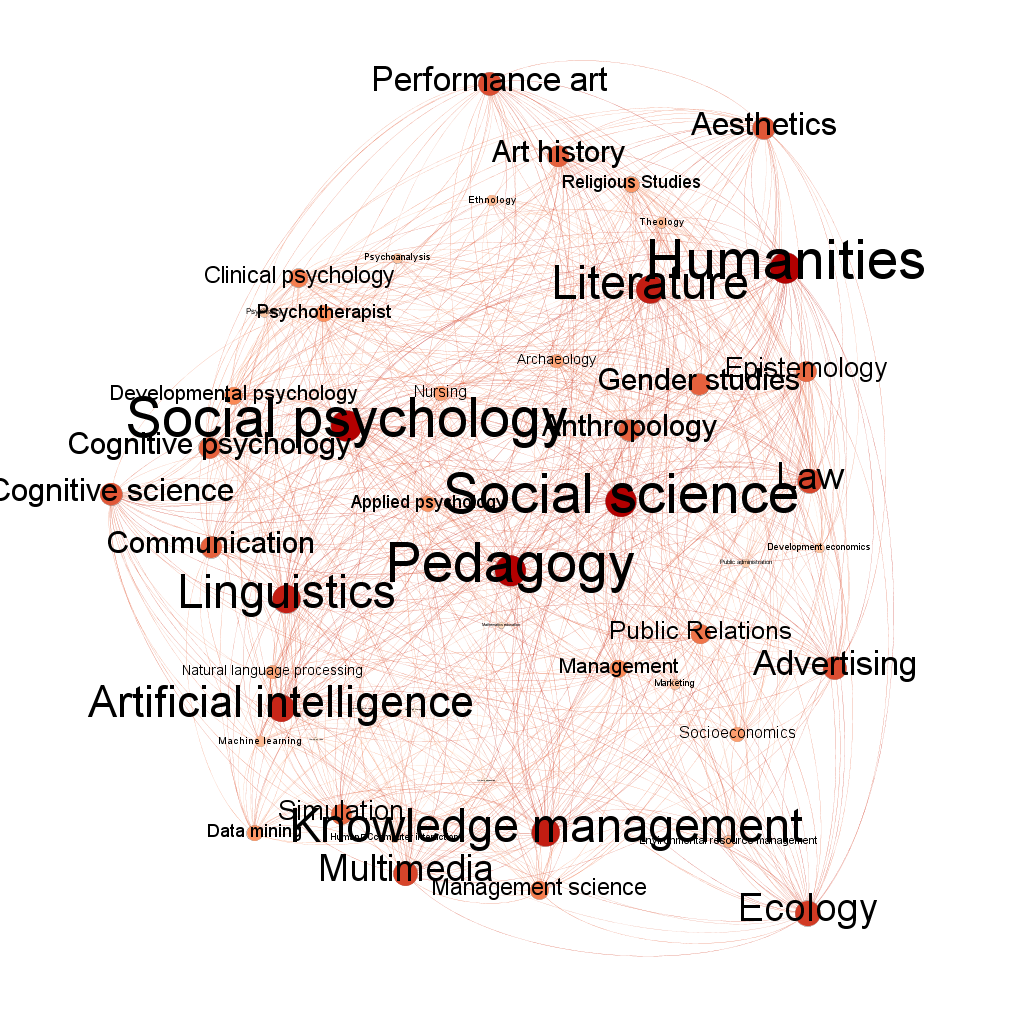}
		\label{fig:12a}
	}%
	\subfigure[Core dataset]{
		\centering
		\includegraphics[width=0.45\textwidth]{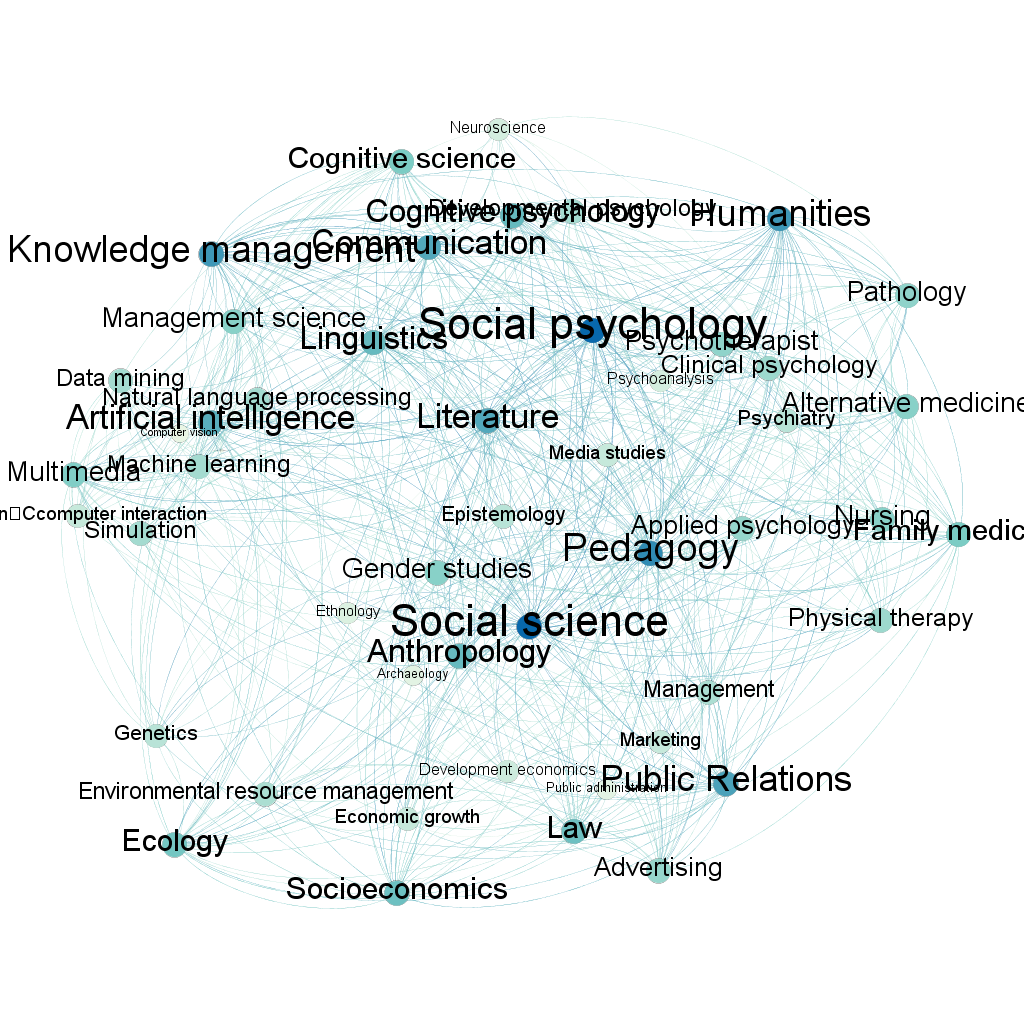}
		\label{fig:12b}
	}%
	\centering
	\caption{Co-presence network of topics.}
	\label{fig:12}
\end{figure}

\subsection{Identifying influential Papers/ Researchers/ Institutions}

Figure \ref{figure10} shows three average self-citation rates: institutional self-citation, paper self-citation, and journal and conference self-citation. In recent years, there has been no obvious growth trend in these areas. As time goes by, scientists are increasingly citing self papers, which may be the reason for the increase in the number of references to a paper.

\begin{table}
	\caption{Ranking of institutions based on the average number of citations per paper from 2000-2017.}
	\label{tab:tabel5}
	\begin{tabular}{lp{3.5cm}<{\centering}p{1.3cm}<{\centering}p{1.3cm}<{\centering}p{1.8cm}<{\centering}p{1.8cm}<{\centering}p{1.8cm}<{\centering}}
		\toprule
		No.&Institution&Number of Researchers& Total Number of Citations&Total Number of Publications&Avg No. of Citations per Paper&Standard Deviation\\
		\midrule
		1&University of California, Berkeley&82&5,636&83&67.90361446&254.5490328\\
		2&Stanford University&69&4,744&75&63.25333333&216.1849242\\
		3&Harvard University&112&3,822&101&37.84158416&77.58578899\\
		4&University of Chicago&52&3,524&51&69.09803922&270.0703801\\
		5&University of Toronto&94&3,042&91&33.42857143&56.17752474\\
		6&University of Melbourne&68&2,823&59&47.84745763&122.0336038\\
		7&McGill University&61&2,776&56&49.57142857&160.2311804\\
		8&University of Wisconsin-Madison&52&2,747&46&59.7173913&199.2495546\\
		9&Princeton University&34&2,691&38&70.81578947&259.6296588\\
		10&Northwestern University&67&2,680&59&45.42372881&92.851235\\
		11&University of British Columbia&85&2,574&77&33.42857143&90.10248968\\
		12&University of California, Los Angeles&70&2,519&57&44.19298246&82.73396232\\
		13&Lancaster University&74&2,433&88&27.64772727&73.89679659\\
		14&University of California, San Diego&65&2,422&63&38.44444444&131.9994841\\
		15&University of Arizona&46&2,261&39&57.97435897&136.5078535\\
		\bottomrule
	\end{tabular}
\end{table}

\begin{figure}[h]
	\centering
	\subfigure[Total dataset]{
		\centering
		\includegraphics[width=0.45\linewidth]{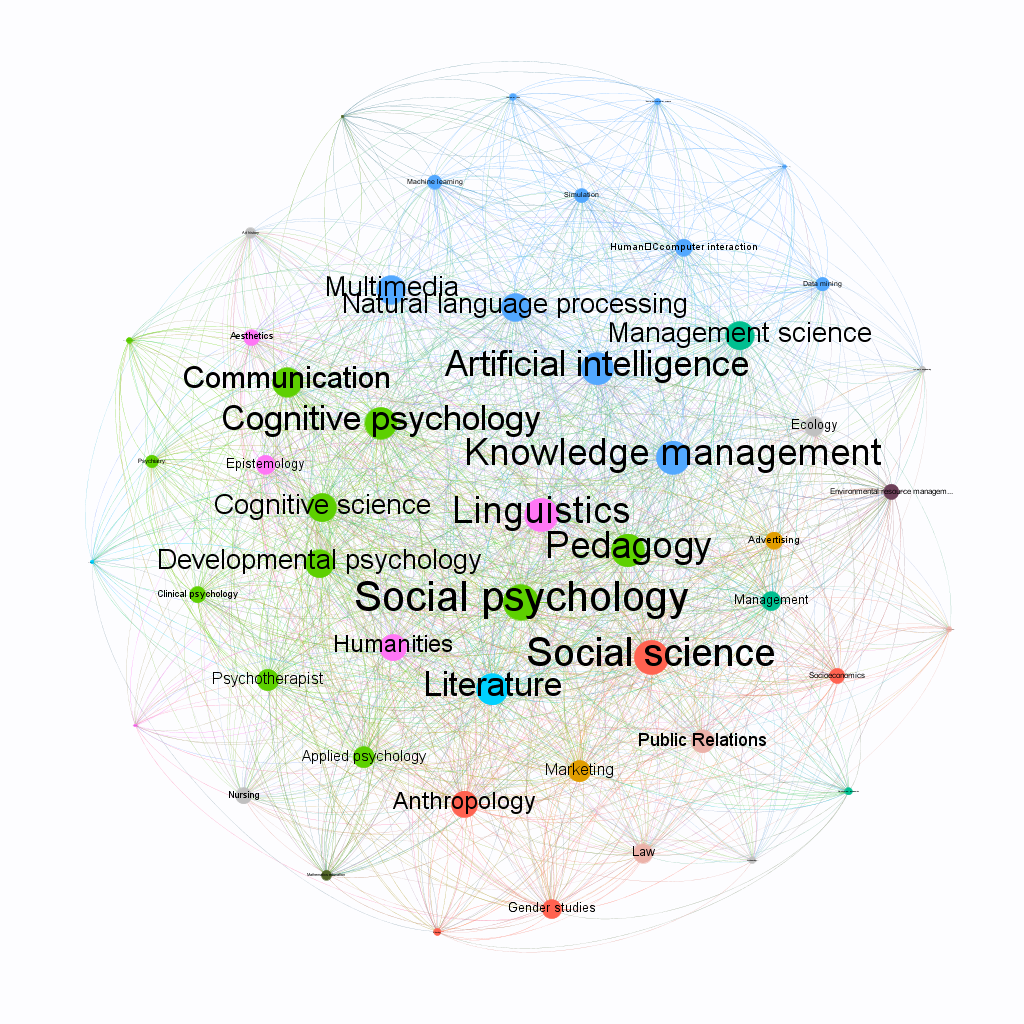}
		\label{fig:13a}
	}%
	\subfigure[Core dataset]{
		\centering
		\includegraphics[width=0.45\linewidth]{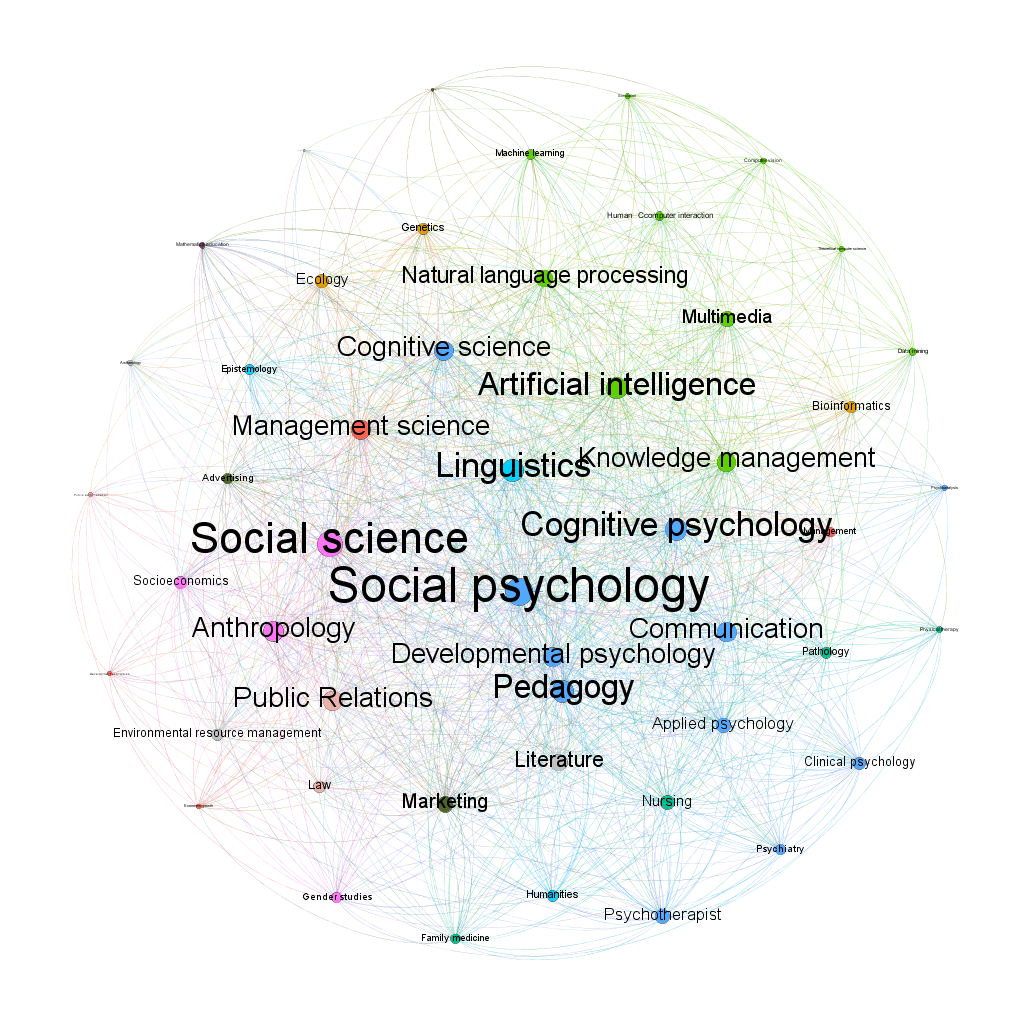}
		\label{fig:13b}
	}%
	\centering
	\caption{Cross-reference network of topics.}
	\label{fig:13}
\end{figure}

We use citations to quantify the importance of paper/res-earcher/institution in metaphor research. For example, we consider the most cited papers from 2000 to 2017 as the most influential papers. Table \ref{tab:table1} and Table \ref{tab:tabel2} show the ranking of papers from 2000 to 2017 based on total citations. By ranking the papers of the two data sets, respectively, the comparison shows that the first eight papers are the same. From the ranking of these papers, we can identify the key issues and keywords in different periods. For example, many social studies in these papers indicate that scholars have invested a lot of time and energy in exploring the relationship between metaphor and society.

Table \ref{tab:tabel3} and Table \ref{tab:tabel4} list the top 15 researchers who cited the most times, as well as the total number of papers they published, the total number of citations, and their affiliations. Although some researchers have published very few papers, they have achieved high citation rates. The quantified top 15 authors with strong influence do not change much between the two data sets.

Research institutions can be seen as clusters of researchers. Table \ref{tab:tabel5} lists the top 15 institutions based on average citations per paper, total number of authors who have published in top journals/conferences, total number of citations, and total number of articles published in top journals/conferences,  in addition to standard deviation of citations (SD) per author and institution. The lower the SD value, the closer the point in the data set is to the average. This can help readers to understand the importance of the target author/institution better. We can see that most of the influential institutions located in North America, Asia, Europe, and Oceania.

Fig. \ref{fig:11} shows the world map embedded with the top 50 most cited institutions and their citation relationships with each other. This can be seen as an overview of citation relationships between influential institutions. According to the citation ranking of papers, influential institutions are distributed in Asia, Europe, North America, and Oceania. As we can see from the figures, citation relations exist widely between North America and Europe. This shows that the dissemination of knowledge is becoming more and more global, and the way it is referenced is also very different. The size of the solid region on the map represents the relative number of agencies cited, and it can be seen that most agencies cited are located in North America. This may be because these institutions get more citations than others.	

\begin{figure}[h]
	\centering
	\includegraphics[width=0.7\linewidth]{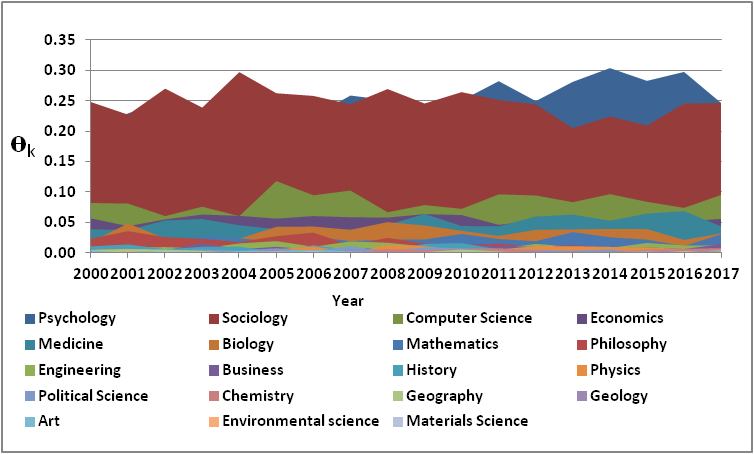}
	\captionsetup{justification=centering}
	\caption{The evolution of core datasets’ topics over time.}
	\label{fig:14}
\end{figure}

\subsection{Internal structure}
Metaphor is not a single topic; it also contains many themes, which are both independent and interactive. To understand metaphor research in-depth, we can divide metaphors into multiple topics by utilizing the existing fields in the dataset. At least the 19 topics with the broadest scope can be divided according to the Level 0 domain.	

Fig. \ref{fig:12} shows the topic co-occurrence network structure as defined in Section \ref{sec2}. Metaphors bring together different topics. For better visualization, we select the top 50 topics with the largest number of papers in the Level-1 field. Fig. \ref{fig:12a} is the topic co-occurrence network of the total data set, which is composed of 779 lines and 50 nodes. Fig. \ref{fig:12b} shows the topic co-occurrence network of the core data set, which is composed of 1,238 lines and 50 nodes. The weight of the lines in the network graph is the coexistence coefficient calculated in the second section, and the degree of topic connection determines the size of the nodes so that the graph can reflect the internal topic structure of metaphor to some extent. The co-occurrence networks of the two data sets are much the same. As shown in the figures, in a paper, it is possible to include topics such as social science, social psychology, and pedagogy. This shows that metaphors contain a variety of topics, their impact, life cycle, and development are different, but they are all interactive. Cross-domain research will promote the continuous development of metaphor.

Also, we apply the methods described above. We divide metaphors into 50 different themes, which are organized by metaphors. Fig. \ref{fig:13} depicts these topics and their references to each other. Unlike the co-occurrence network mentioned above, the weights of lines in the cross-reference network are measured according to the number of papers on the cited topic. Nodes of the same color belong to the same Level-0 field. For example, in Fig. \ref{fig:14}, the green nodes consist of the sub-fields of computer science in the Level-0, such as natural language processing, artificial intelligence, and multimedia, and the blue nodes comprise the sub-field of psychology such as social psychology, pedagogy, and cognitive psychology. Through the connections of different color nodes, it is easy to see that metaphor research is cross-domain rather than independent.

In addition, to reorganize the topic dynamically, as defined in the previous section, we took $\theta_k^{[t]}$ over the evolution of topic $k$. Fig. \ref{fig:14} shows the topic change over time in 19 domains at Level 0 from 2000 to 2017. These topics are ranked from bottom to top in terms of popularity. From the core data sets, it can be seen that metaphor research focuses more on topics such as psychology and sociology, and it pays less attention to environmental science, materials science, and other topics. This figure also clearly reflects the evolution of the topics, some of which have been declining over time, while others have received much attention.

To investigate the popularity of topics further, we use the $r_k$ defined in Section \ref{sec2} to evaluate these topics. Table \ref{tab:tabel6} lists $r_k$ estimates for all topics in descending order. The hottest topics are chemistry, business, and physics.

\begin{table}
	\caption{Increase index for popular topics.}
	\label{tab:tabel6}
	\begin{tabular}{p{3cm}<{\centering}p{1cm}<{\centering}p{3cm}<{\centering}p{1cm}<{\centering}p{3cm}<{\centering}p{1cm}<{\centering}}
		\toprule
		Topic & rk & Topic & rk & Topic & rk \\
		\midrule
		Chemistry & 2.02  & Medicine & 0.96  & Geography & 0.77 \\
		Business & 1.10  & Mathematics & 0.94  & Economics & 0.74 \\
		Physics & 1.07  & Computer Science & 0.85  & Sociology & 0.74 \\
		Geology & 0.99  & Political Science & 0.82  & Materials Science & 0.73\\
		Art & 0.30 & Philosophy & 0.38 & History & 0.58 \\
		Environmental Science & 0.69 &&&&\\
		\bottomrule
	\end{tabular}
\end{table}

\section{Conclusion}
\label{sec4}
In this paper, we undertook a bibliographic analysis of metaphor research in the 21\ts{st} century. To reflect the universality of the law, we took 11,564 articles from the Web of Science as the core data set, and 44,586 papers from MAG as the whole data set. We perform the same calculations and compare the results of the two data sets. We conduct statistical analyses of the titles, authors, institutions, and reference data of each paper. We also provide a relatively comprehensive review of metaphor development over the past 18 years.

We found that the results of the two data sets are roughly the same. From the perspective of publications, authors, citations, and references, metaphor research generally shows an upward trend. From the perspective of changes in reference behavior, the development trend of metaphor is open and popular, which is reflected over time. The number of references is increasing, and cross-domain metaphor research is becoming more and more common. From the changes in the number of citations and publications, we observe that the trend of cooperation is becoming more and more obvious, and the average productivity of each researcher is declining. To quantify the development of metaphor studies better, we use the average number of citations of each paper per author/author/institution as an indicator of its importance, ranking the importance of the paper/author/institution, and screening out excellent papers, authors, and countries in the field of metaphor research. Finally, we explore the internal structure of metaphors, and we conclude that the field contains a variety of complex and changing themes, with differences and connections between them. These findings reveal the evolution of potential patterns and themes in the metaphorical world, helping researchers  continue to explore the field and providing them with novel insights.

\section*{Acknowledgment}
This work is partially supported by National Natural Science Foundation of China under Grants No. 62076051. We would like to thank Wei Zhang for help with experiments.

\bibliographystyle{splncs03}
\bibliography{Metaphor}


%
%


\end{document}